\documentclass[final,5p,times,twocolumn]{elsarticle}

\usepackage{hyperref}
 \usepackage{graphicx}

\usepackage{amssymb}
\usepackage{amsthm}

\usepackage[nodots,nocompress]{numcompress}

\biboptions{}

\journal{Computer Physics Communications}

\begin{document}

\begin{frontmatter}


\title{Java Application for Superposition T-matrix Code to Study the Optical Properties of Cosmic Dust Aggregates}


\author[rvt]{P. ~Halder}
\ead{prithishh3@gmail.com}
\author[rvt]{A. ~Chakraborty}
\ead{carindam1@gmail.com}
\author[rvt]{P. ~Debroy}
\ead{pari.hkd@gmail.com}
\author[rvt]{H. S. ~Das\corref{cor1}}
\ead{hsdas@iucaa.ernet.in}
\address[rvt]{Department of Physics, Assam University, Silchar 788011, India}
\cortext[cor1]{Corresponding author}

\begin{abstract}
In this paper, we report the development of a java application for the Superposition T-matrix code, JaSTA (Java Superposition T-matrix App), to study the light scattering properties of aggregate structures. It has been developed using Netbeans 7.1.2, which is a java integrated development environment (IDE). The JaSTA uses double precession superposition codes for multi-sphere clusters in random orientation developed by Mackowski \& Mischenko (1996). It consists of a graphical user interface (GUI) in the front hand and a database of related data in the back hand. Both the interactive GUI and database package directly enable a user to model by self-monitoring respective input parameters (namely, wavelength, complex refractive indices, grain size, etc.) to study the related optical properties of cosmic dust (namely, extinction, polarization, etc.) instantly, i.e., with zero computational time. This increases the efficiency of the user. The database of JaSTA is now created  for a few sets of input parameters with a plan to create a large database in future. This application also has an option where users can compile and run the scattering code directly for aggregates in GUI environment. The JaSTA aims to provide convenient and quicker data analysis of the optical properties which can be used in different fields like planetary science, atmospheric science, nano science, etc. The current version of this software is developed for the Linux and Windows platform to study the light scattering properties of small aggregates which will be extended for larger aggregates using parallel codes in future.

\vspace{0.8cm}
\hspace{-0.4cm}
\textbf{Program Summary:}

\begin{small}
\noindent
{\em Program title:} JaSTA: Java Superposition T-matrix App.                                          \\
{\em Catalogue identifier:} AETB\_v1\_0                                    \\
{\em Program summary URL:} \url{http://cpc.cs.qub.ac.uk/summaries/AETB\_v1\_0.html}   \\
{\em Program obtainable from:} CPC Program Library, Queen’s University, Belfast, N. Ireland.   \\
{\em Licensing provisions:} Standard CPC licence, \url{http://cpc.cs.qub.ac.uk/licence/licence.html}  \\
{\em No. of lines in distributed program, including test data, etc.:} 571570.         \\
{\em No. of bytes in distributed program:} 120226886. \\
{\em Distribution format:} tar.gz \\
{\em Programming language:}  Java, Fortran95.                                    \\
{\em Computer:} Any Windows or Linux systems having java runtime environment, java3D and fortran95 compiler; Developed on 2.40 GHz Intel Core i3.                                               \\
{\em Operating system:} Any Windows or Linux system having java runtime environment, java3D and fortran95 compiler.                                       \\
{\em RAM:} Ranging from few Mbytes to several Gbytes, depending on the input parameters.                                              \\
{\em Classification:} 1.3 Radiative Transfer.                                         \\
{\em External routines/libraries:} jfreechart-1.0.14 \cite {JFreeChart 2012} (free plotting library for java), j3d-jre-1.5.2 \cite {java3D 2013} (3D visualization).                            \\

\end{small}
\end{abstract}


\end{frontmatter}


\section{Introduction}
Dust plays an important role in the overall scenario of structure and evolution of the universe. Cosmic dust is a kind of dust that is composed of particles in space which are irregular in shape whose porosity ranges from fluffy to compact. Cosmic dust includes comet dust, asteroidal dust, interstellar dust and interplanetary dust. Extensive and high quality observations carried out using ground-based telescopes, satellites and theoretical calculation based on different mathematical models have led to phenomenal progress in our understanding of the nature and composition of dust in the universe. These studies show that dust grains are effective absorbents and scatterer of electromagnetic radiation energy.
The interplanetary dust particles (IDPs) which are collected from the Earth's stratosphere by high-flying aircraft \cite{Brownlee 1985}, \cite{Warren 1994} usually have irregular shapes and fluffy structures. Similar structures have been produced in laboratory and microgravity experiments of dust particle interactions \cite{Wurm 1998}, \cite{Blum 2000}, \cite{Krause 2004}. It has also been suggested that interstellar dust grains may consist primarily of such aggregate structures \cite{Mathis 1989}, \cite{Dorschner 1995}, with a mixture of various chemical compositions and vacuum. It is now well accepted from the in situ measurements of comets and Stardust-returned samples of Comet Wild 2 that cometary dust consists of a mixture of compact particles and aggregates \cite{Lamy 1987}, \cite{Fomenkova 1999}, \cite{Horz 2006}, \cite{Zolensky 2006}, \cite{Burchell 2008}, etc.

Dust particles present in comet scatter and absorb the incident solar radiation. Our knowledge of cometary dust comes from polarimetric studies of comets, remote observation of infrared spectral features and the in situ measurement of comets (e.g. comet Halley and the Stardust mission). The polarization measurement of the scattered radiation gives valuable information about the shape, structure and sizes of the dust particles. Many investigators \cite{Kikuchi 1987},  \cite{Mukai 1987}, \cite{Sen 1991a}, \cite{Sen 1991b}, \cite{Xing 1997}, \cite{Petrova 2000}, \cite{Das 2004}, \cite{Petrova 2004}, \cite{Bertini 2007}, \cite{Kolokolova 2007},  etc. have studied linear and circular polarization measurements of several comets. These studies enriched the knowledge about the dust grain nature of comets.
Also modeling of the wavelength dependencies of interstellar extinction and linear polarization allows us to obtain different information about geometrical properties (viz. size, shape, etc.), abundance of elements and composition of the interstellar dust. Interstellar polarization also tells us about the structure of the magnetic fields due to dichroic extinction of non-spherical grains aligned in large-scale Galactic magnetic fields \cite{Greenberg 1968}, \cite{ Hough 2004}.

Optical properties of aggregated particles have been extensively investigated through the use of various numerical techniques. The Superposition T-matrix (STM) code \cite{Mackowski 1996} and the Discrete-Dipole Approximation (DDA) code \cite{Purcell 1973}, \cite{Draine 1988}, \cite{Draine 1994} are widely used by researchers to study the light scattering properties of cosmic dust aggregates. The T-matrix approach involves a superposition solution to Maxwell's equations for the multiple spherical boundary domain. In this technique, the scattered field from the ensemble of N spheres is represented by the superposition of fields scattered from each of the spheres in the ensemble. Each of the individual sphere (or partial fields) is represented by an outgoing wave vector spherical harmonic expansion, centered about the origin of sphere. The main advantage of the T-matrix formalism is the orientation average of scattering matrix that can be performed analytically. Therefore, calculation with the STM code is fast. The DDA code is used to calculate scattering and absorption of electromagnetic waves by targets with arbitrary geometries. In this approximation the target is replaced by an array of N point dipoles (or polarizable points), with the spacing between the dipoles small comparable to the wavelength; the electromagnetic scattering problem for an incident periodic wave interacting with this array of point dipoles is then solved exactly. We presently use the STM code in our software package as its computation time is much less compared to that of the DDA code. The T-Matrix method for calculating the light scattering by nonspherical particles,  based on numerically solving Maxwell's equation was developed by Peter Waterman \cite{Waterman 1965}, \cite{Waterman 1971} as a technique for computing electromagnetic scattering by single, homogeneous non-spherical particles based on the Huygens principle.  At present, the T-matrix approach is one of the most powerful and widely used tools for rigorously computing electromagnetic scattering by single and compounded non-spherical particles. It compares favorably with other frequently used techniques in many applications  in terms of efficiency, accuracy, and size parameter range. This is the only method that has been used in systematic surveys of non-spherical scattering based on calculations for thousand of particles in random orientation.

The T-matrix theory was originally developed for homogeneous star-shaped particles with axisymmetric \cite{Waterman 1965}, \cite{Mishchenko 1994} and non-axisymmetric \cite{Laitinen 1998}, \cite{Kahnert 2005} geometries. It has been developed in the case of ensembles of spheres, i.e., aggregated particles by means of a superposition approach \cite{Mackowski}. The STM approach involves a superposition solution to Maxwell's equations for the multiple spherical boundary domain where one can compute the light scattering properties of aggregate particles in either fixed or random orientations. This theory was extensively used by astronomers to study the optical properties of cosmic dust aggregates (\cite{Kimura 2003}, \cite{Kimura 2006}, \cite{Kolokolova 2007}, \cite{Bertini 2007}, \cite{Shen 2008}, \cite{Shen 2009}, \cite{Lasue 2009}, \cite{Paul 2010}, \cite{Das 2010}, \cite{Das 2011}, etc).

The STM theory is also used to study optical properties of atmospheric aerosols and nanoparticles. Soot particles are an important factor in regulating climate and weather process. Liu et al. \cite{Liu 2008} studied scattering and absorption properties of soot aggregates with varying state of compactness and size using the STM method. They found that the fractal dimension is an important parameter for the evaluation of the optical properties of a soot cluster. Recently, Takano et al. \cite{Takano 2013}  studied the single-scattering properties of black carbon soot aggregates using geometric-optics surface-wave (GOS) approach and compared their results with those determined from the STM method. They found that under the random orientation condition, the single-scattering results determined from GOS compare reasonably good with those obtained from STM. Messina et al. \cite{Messina 2011} studied how light forces can be used to trap gold nanoaggregates of selected structure and optical properties obtained by laser ablation in liquid. Using T-matrix formalism of light scattering theory to the optical trapping of metal aggregates, they showed how the plasmon resonances and aggregate structure are responsible for the increased forces and wider trapping size range with respect to individual metal nanoparticles.

In the present work, we develop graphical user interface (GUI) platform of STM code to study the light scattering properties of aggregate structures. In order to perform  light scattering calculations in command line environment, one has to keep in mind several details ,e.g., it is necessary to create specific configuration files in specific folders and  paths; within these configuration files the corresponding light scattering parameters have to be named and listed in a specific way, etc. These are time-consuming tasks and additionally can lead to errors. All these complexities can discourage a potential user from trying and using the available STM code for their scientific purposes. To cope with this common problem in scientific programming, graphical user interfaces (GUI) version of such programs are highly beneficial over application programming interface (API).  Although many command line environments are capable of multitasking, they do not offer the same ease and ability to view multiple things at once on one screen. GUI users have windows that enable a user to view, control, and manipulate multiple things at once and is much faster to navigate when compared with a command line. The graphical representation of the results within a single package will help the user to analyze and compare the results obtained during computations. These results can be saved and used in ongoing or forthcoming research. On the other hand the database of the software package also saves lots of computation time of the researcher, generally required in performing live computations. This makes the research much easier and error free. So the user from different branches of science can use GUI-based application to continue their research work without concentrating much on complex sequential steps of programming to get the results.
\section{Aggregate Dust Model}
The studies of interplanetary and cometary dust indicate that cosmic grains are likely to be porous, fluffy and composites of many small grains coalesced together, due to grain-grain collisions, dust-gas interactions and various other processes \cite{Krueger 1989}, \cite{Greenberg 1990}, \cite{Wolff 1994}. Porous, composite aggregates are often modelled as a cluster of small spheres (``monomers"), assembled under various aggregation rules with typical sizes 0.1-10$\mu m$. Here grain aggregates are assumed to be fluffy sub-structured collections of minute particles which are loosely held to one another. Each particle is assumed to consist of a single material, such as silicates or carbon, as formed in the various separate sources of cosmic dust.
To study natural aggregates, several investigators \cite{Kimura 2003}, \cite{Kimura 2006}, \cite{Kolokolova 2007}, \cite{Bertini 2007}, \cite{Das 2008a}, \cite{Das 2008b}, \cite{Das 2010}, \cite{Das 2011}, \cite{Lasue 2009} etc. built aggregates using ballistic aggregation procedure \cite{Meakin 1983}, \cite{Meakin 1984}. Using Monte-Carlo simulation, these aggregates are built by random hitting and sticking particles together. In the case when the procedure allows only single particle to join the cluster of particles, the aggregate is called ballistic particle-cluster aggregate (BPCA), if the procedure allows clusters of particles to stick together, the aggregate is called ballistic cluster-cluster aggregate (BCCA). Usually BCCA are more porous, whereas BPCA have more compact arrangement of particles. The BPCA and  BCCA structures generated by our code are provided with the JaSTA package. Recently \cite{Shen 2008}, \cite{Shen 2009} constructed aggregates using three specific aggregation rules: ballistic agglomeration (BA), ballistic agglomeration with one migration (BAM1) and ballistic agglomeration with two migrations (BAM2) to characterize the irregular shapes of the aggregates. The newly-introduced BAM1 and BAM2 clusters have geometries that are random but substantially less ``fluffy'' than the BA clusters.
\begin{figure}
\centering
\includegraphics[width=75mm]{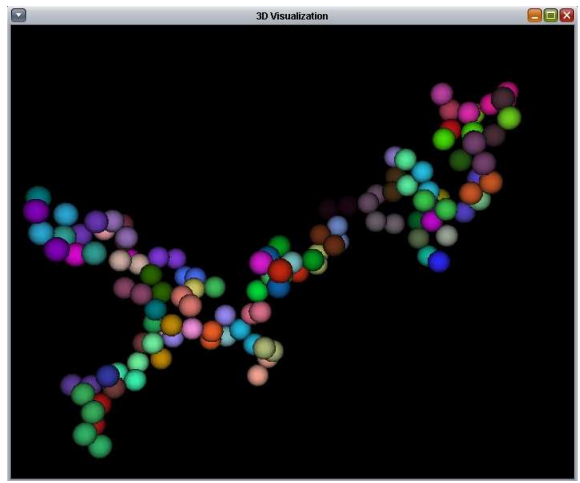}
\caption{3D visualization of a BCCA structure for 128 monomers.}

\end{figure}
\begin{figure}
\centering
\includegraphics[width=75mm]{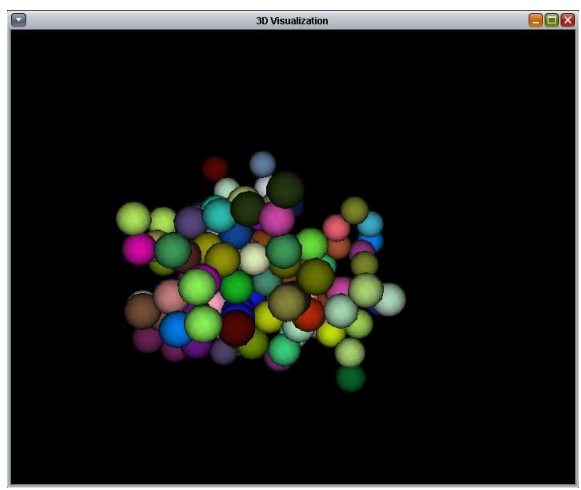}
\caption{3D visualization of a BPCA structure for 128 monomers.}
\end{figure}
\subsection{3D Visualization}
We developed a code to generate the BPCA and BCCA structures which is based on Monte-Carlo simulations. This code generates x, y and z position coordinates of the structures for monomers of radius 1.  The aggregates are created for  number of monomers, N = 8, 16, 32, 64 and 128. In order to get the 3D feel and look of the structure, we have incorporated java3D into JaSTA. We have designed our own codes using Java 3D library to create an array of virtual spheres in a virtual 3D space whose positions are guided by the coordinates of the structure file. Fig-1 and Fig-2 show the 3D visualization of  BCCA and  BPCA structures of 128 monomers respectively. This 3D structure can be rotated by dragging the mouse horizontally or vertically over the 3D structure. By scrolling the mouse wheel over the 3D structure one can zoom in or zoom out the structure. More details and video tutorials are present in our website \verb"ausastro.in".

\section{Java Superposition T-matrix App (JaSTA)}
Java Superposition T-matrix App (JaSTA) is a java swing application aimed to study the light scattering properties of cosmic dust aggregates based on Mackowski \& Mischenko's STM code. The application software is developed using Netbeans 7.1.2, which is a J2SE integrated development environment (IDE) available for Windows, Linux, Mac OS and Solaris. Netbeans is an open source project dedicated to provide software development products that addresses the need of developers and users.  It is one of the most commonly used IDE for developing cross-platform desktop applications using java. It has built-in tools to design the graphical user interface which can be easily connected to any algorithm for background calculation. Due to such flexible utility of Netbeans we have chosen it to design and develop JaSTA.  JaSTA generates 3D visualization of various aggregate structures using Java3D technology. Java3D is a 3D application programming interface for the java platform. It runs either on OpenGL or Direct 3D. JaSTA also uses JFreeChart-1.0.14 java library in order to plot various graphs. The present version of the software package is JaSTA-1.0.1 and is available for download from the website \url{http://ausastro.in/jasta.html}. A user manual JaSTA Manual-1.0 and a video tutorial for the current version of the software are also available in the above-mentioned web link.

\section{Input and Output parameters of JaSTA}
The list of input parameters of JaSTA is:
\begin{enumerate}
\item{radius of monomer ($a_m$)},
\item{wavelength ($\lambda$) of incident radiation},
\item{real refractive index ($n$)},
\item{imaginary refractive index ($k$)},
\item{number of monomers ($N$)} and
\item{aggregate type (BCCA, BPCA \& user$\_$structure)}.
\end{enumerate}

The list of output parameters of JaSTA is:
\begin{enumerate}

\item{the extinction efficiency} (Q$_{ext})$,
\item{the absorption efficiency} (Q$_{abs})$,
\item{the scattering efficiency}  (Q$_{sca})$,
\item{extinction cross section}    (C$_{ext})$,
\item{absorption cross section}    (C$_{abs})$,
\item{scattering cross section}    (C$_{sca})$,
\item{albedo} ($\varpi$),
\item{a dimensionless size parameter (xscale = $\frac{2 \pi a}{ \lambda}$), where $a$ is the radius of the monomers,  and $\lambda$ is the incident wavelength } and
\item{the nonzero  scattering matrix elements viz. ``$S11$", ``$-(S12/S11)$", ``$S33/S11$" and ``$S34/S11$" which are functions of scattering angle.}
\end{enumerate}

The detailed description of input and output parameters are discussed in Mackowski \& Mischenko {\cite{Mackowski 1996}} and the manual provided with their STM code (``scsmfo.ps" file).

\begin{figure}
\centering
\includegraphics[width=74mm]{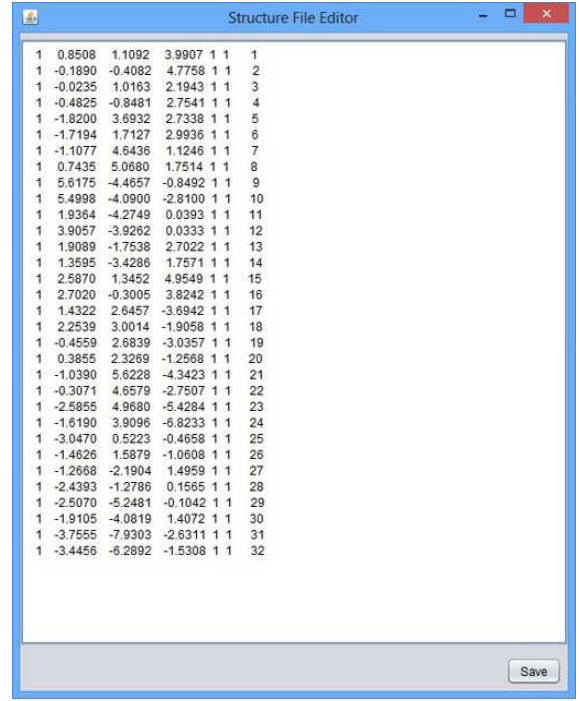}
\caption{Structure file editor.}
\end{figure}

\section{Structure File}
JaSTA provides a unique feature of user-defined structure file to the calculation. Apart from the default BCCA and BPCA structures there is another option for the user to provide their own structure to the application. The format of the structure is shown in Fig-3. To study the light scattering properties of cosmic aggregates, we take BPCA and BCCA structures. The BCCA and BPCA structures have been generated using Monte - Carlo simulation. Similarly the user can generate a structure file using a different algorithm and edit the user structure file in the JaSTA package and then proceed for calculation. On clicking the user structure option in the aggregate type combobox, a window appears in order to edit the structures or coordinates.

\begin{figure*}[!ht]
\centering
\includegraphics[width=90mm]{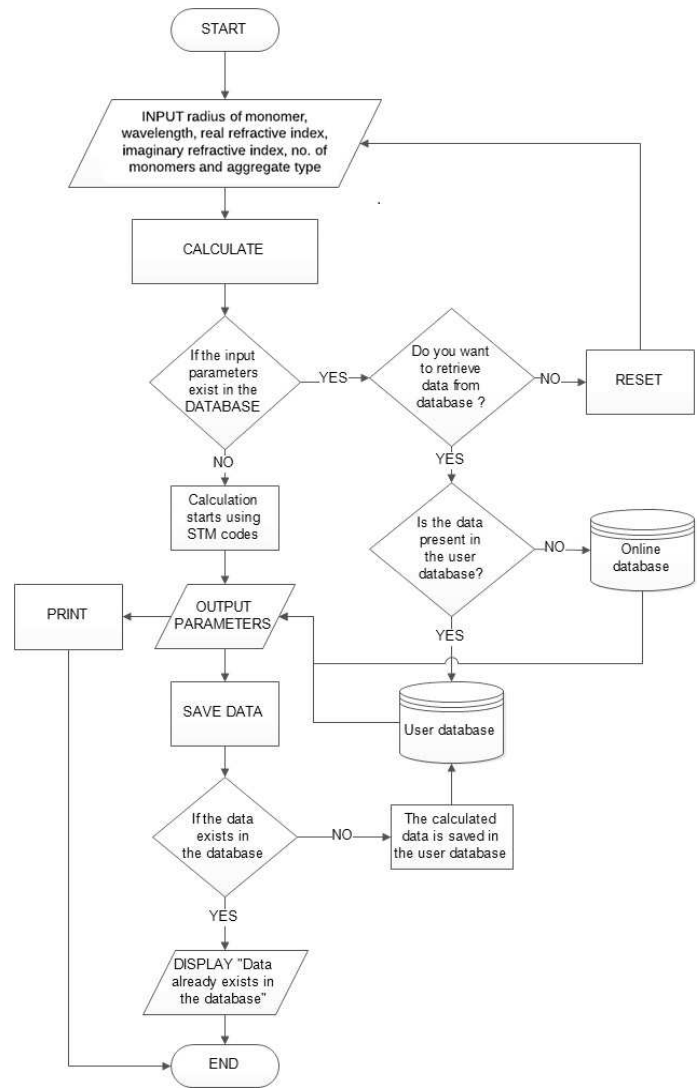}
\caption{Flow chart of JaSTA.}
\end{figure*}

\begin{figure*}[!ht]
\centering
\includegraphics[width=120mm]{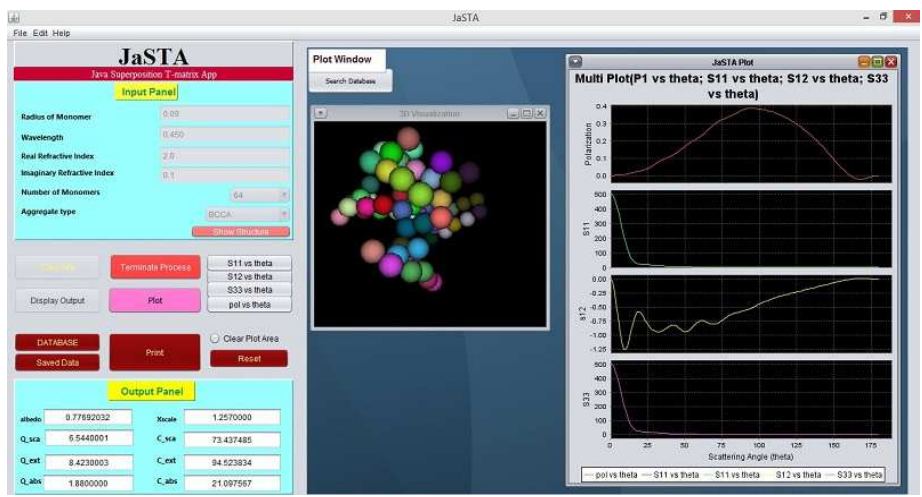}
\caption{Home window of JaSTA.}
\end{figure*}

\section{Details of the Application}
JaSTA consists of a graphical user interface (GUI) in  the front hand and a database of related data in the back hand. The flowchart of the application shown in Fig-4 explains the working structure of JaSTA. It takes input parameters from the user, searches the input data in the database, if the data is present in the database the output parameters are called from the database and displayed on the screen, if the data is absent in the database, the STM codes start calculation in accordance with the input data and displays the calculated output parameters on the screen. The main window of the GUI is shown in Fig-5. The application consists of four main sections. The input panel, control panel, output panel and the graph area.
 Fig-6. shows the input panel, where we have considered monomer radius ($a_m$) = 0.09 $\mu{m}$, wavelength of incident radiation ($\lambda$) = 0.450 $\mu{m}$, real refractive index ($n$) = 2.0, imaginary refractive index ($k$) = 0.1, number of monomers (N) = 64 and aggregate type = BCCA (Ballistic Cluster Cluster Aggregation). One can select a number of monomers from 8 to 128 in the corresponding combobox. The user can also choose aggregate type: BCCA, BPCA and user\_structure (user-defined structure file). The 3D visualization of the structure can be viewed by clicking the ``Show Structure" button in the input panel. The 3D visualization is discussed in section 2.1.
  \begin{figure}[!ht]
  \centering
\includegraphics[width=70mm]{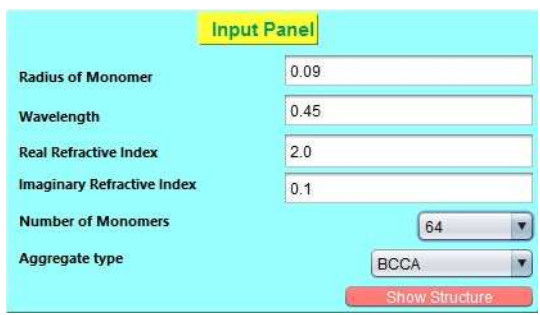}
\caption{Input Panel of JaSTA.}
\end{figure}
 \begin{figure}[!ht]
 \centering
\includegraphics[width=70mm]{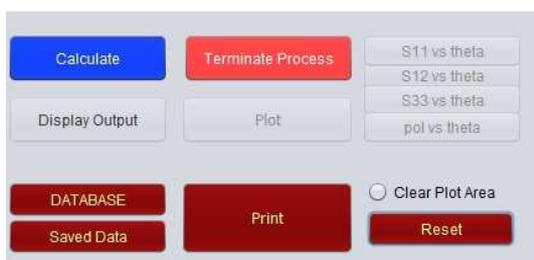}
  \caption{Control Panel of JaSTA.}

\end{figure}
\begin{figure}[!ht]
\centering
\includegraphics[width=70mm]{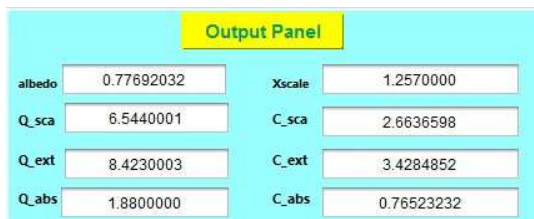}
\caption{Output Panel of JaSTA.}
\end{figure}
Fig-7. portrays the control panel, from where all the actions like, live calculation of  data entered in the input panel, calling pre-calculated data from the database, plotting different graphs and clearing a particular calculation can be controlled. Calculation button performs an instant database check of the data  entered in the input panel. If the data is found in the database then the corresponding result is shown in the output panel as output parameters and in the plot area as plots of various scattering co-efficient vs. scattering angle. If the entered data does not exist in the database, live calculation starts i.e., the superposition T-matrix code  starts its operation in accordance with the input data. During live calculation a progress bar and a console window appear. The progress bar shows the percentage of completion for calculation and the console window shows the details of each matrix element with time.
\begin{figure}
\centering
\includegraphics[width=70mm]{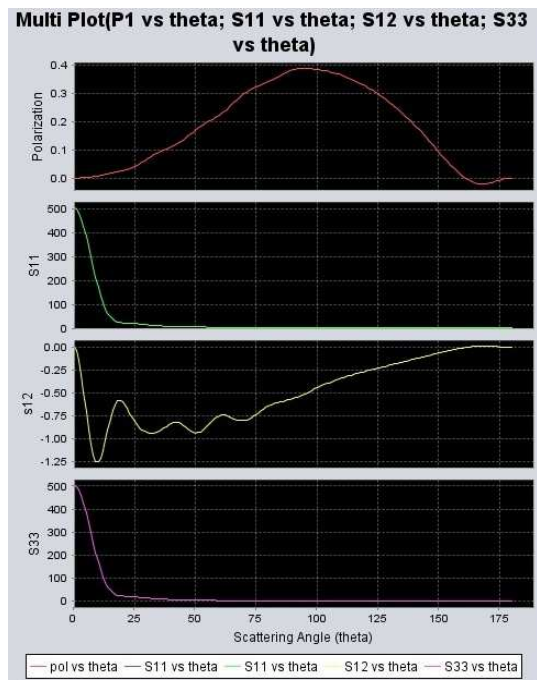}
\caption{Multiplot window showing variation of various scattering coefficients with the scattering angle.}
\end{figure}
 After the completion of calculation one can retrieve the output parameters by clicking the ``Display Output" button. Fig-8. shows the output panel with output parameters generated by the STM codes for the above input parameters. The output parameters are (i) the extinction efficiency factor (Q$_{ext}$), (ii) the absorption efficiency factor (Q$_{abs})$, (iii) the scattered efficiency factor (Q$_{sca})$, (iv) the extinction cross section  (C$_{ext})$, (v) the absorption cross section  (C$_{abs})$, (vi) the scattering cross section (C$_{sca})$, (vii) albedo ($\varpi$) and (viii) size parameter of monomer (xscale). The ``Plot" button appears once the ``Display Output" button is clicked in the control panel. It generates a multi-plot window in the plot area for scattering matrix elements viz. ``S11", ``$-$(S12/S11)", ``S33/S11" and ``S34/S11" which are functions of scattering angle as shown in Fig-9. The entire calculation for a particular set of input parameters can be saved to the user database by the ``Save" option in the file menu. The user data is saved in the folder ``DATA" in the installation folder of JaSTA which constitutes the user database. The user database and the online database are discussed in the next section.
\begin{figure}
\centering
\includegraphics[width=100mm]{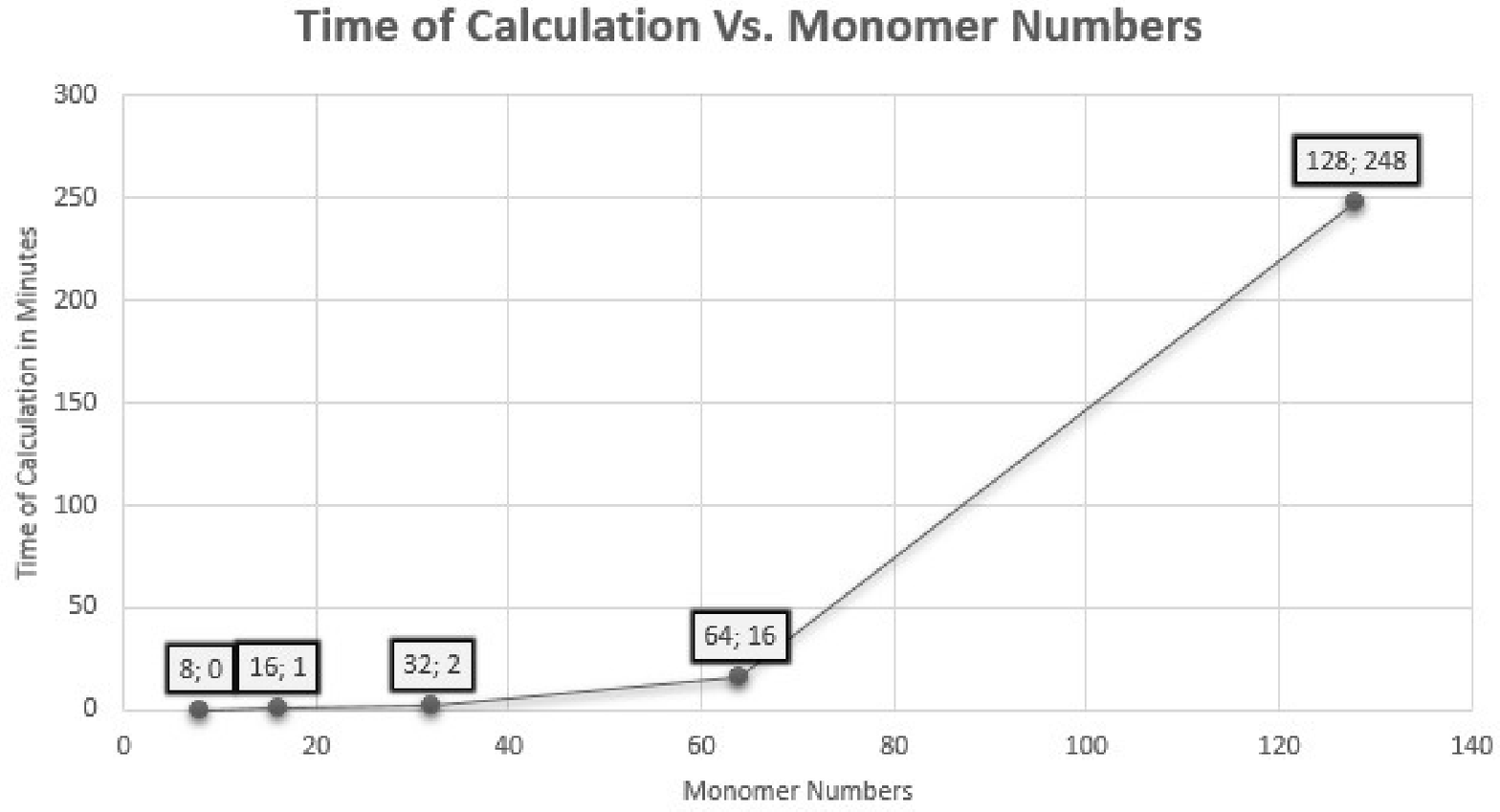}
\vspace{1cm}
\caption{Variation in calculation time with increasing number of monomers.}
\end{figure}

\begin{figure*}
\centering
\includegraphics[width=130mm]{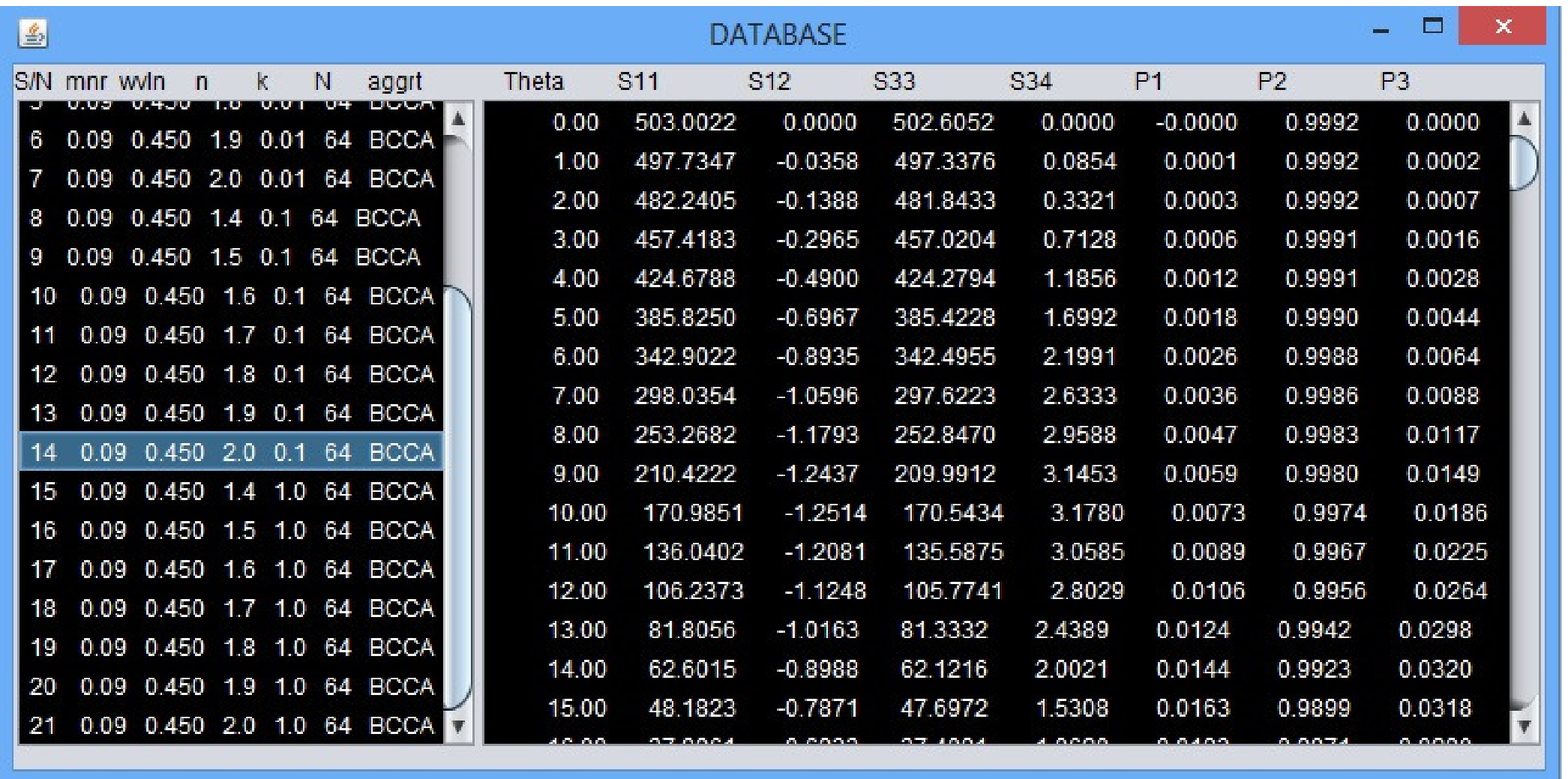}
\caption{Database of saved data.}
\end{figure*}

\section{Database}
The computation time increases when the number of monomers in aggregate increases. Fig-10 shows the variations in calculation time with increasing number of monomers. So the need of enormous amount of pre-calculated data emerges for better and quicker access of data analysis. Hence we have provided an advanced database feature in JaSTA.

The database is divided into three sections, (i) User database, (ii) Data Files and (iii) Online database.
\subsection{User database :}
It is the database of calculated data saved by the user and it can be accessed using the ``Saved Data" button. Fig-11 shows the user database window with two columns. The left column contains the input parameters of the saved data with a particular serial number. On clicking a particular set of input data in the left panel, the corresponding scattering matrix elements appear in the right panel and the corresponding output parameters and plots appear in the output panel and the plot area respectively.

This database contains data provided with JaSTA, calculated in the High Computational Laboratory of Department of Physics, Assam University, Silchar (India). It can be accessed by clicking the ``Database" button. The structure and working of the database are similar to the user database. We are also planning to provide automatic update of the online database in the future versions of JaSTA.

\subsection{Data Files:}

The database of JaSTA contains data files which can be identified from their unique name. This name is assigned in a general format for all data files so that the user can identify the input parameters directly from the name of the data file. The format is shown in Fig-12. The format can be decoded from the characters present in the file name from left to right i.e., $(i)$ serial number, $(ii)$ radius of monomer, $(iii)$ wavelength, $(iv)$ real refractive index $(v)$ imaginary refractive index, $(vi)$ number of monomers and $(vii)$ aggregate type.
\begin{figure}
\centering
\includegraphics[width=74mm]{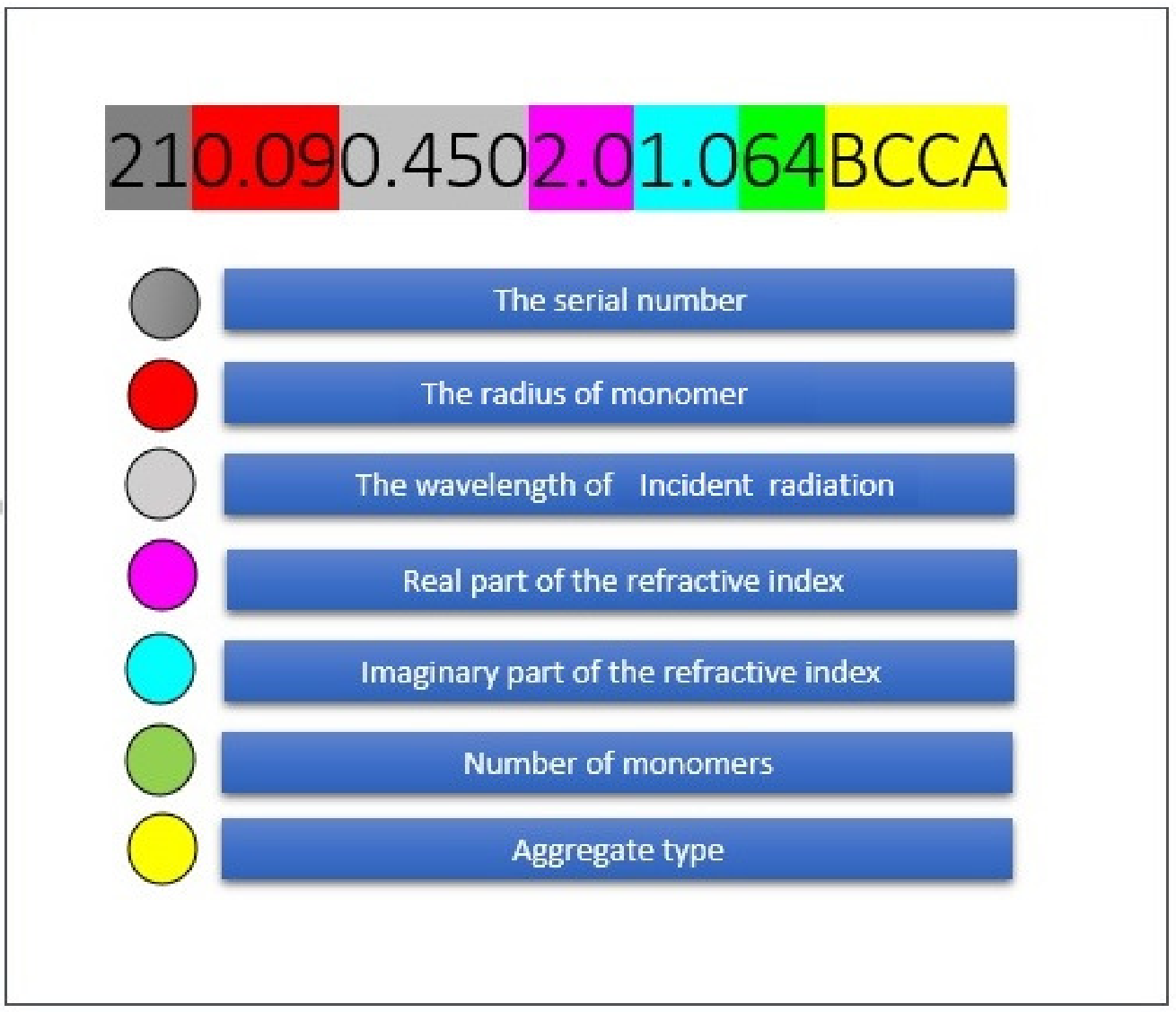}
\caption{General name format for a data file.}
\end{figure}

\begin{figure*}[!ht]
\centering
\includegraphics[width=110mm]{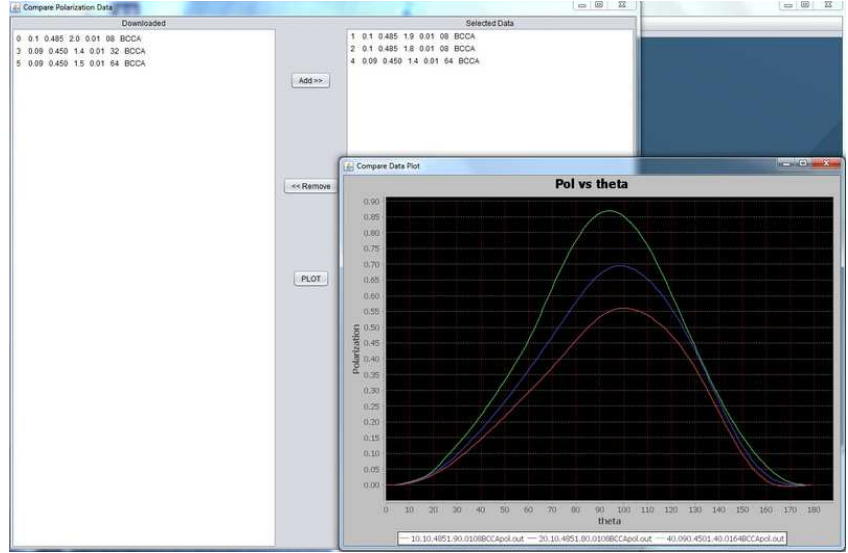}
\caption{Comparison of plots generated from three different set of input parameters.}
\end{figure*}
\subsection{Online database:}
This database contains data provided with JaSTA, calculated in the High Computational Laboratory of Department of Physics, Assam University. It can be accessed by clicking the ``Database" button. The structure and working of the database are similar to the user database. The online database will receive periodic and automatic updates from our server.
\section{Feature of JaSTA Package}

For more analytical and non-trivial solution of various parameters the compare plot option has been introduced. This version of the software will have analytical approach only for polarization values. One can select any set of data from the database and compare their polarization values with help of ``Compare Plot" option as shown in Fig-13. We have shown the change in polarization with the variation in real refractive indices (Fig-14), monomer number (Fig-15), monomer size (Fig-16) and aggregate type (Fig-17) . This option reduces the use of external plotting applications like `Gnuplot', `Origin', etc., for comparison of various sets of data. We also plan to introduce new option for comparison of numerical data with observational data in future.

\section{Data Analysis}

The data analysis is performed using the ``Compare Plot" option. We have calculated a large number of data by varying different parameters like monomer number, monomer size, refractive indices and the aggregate type and saved the results in the ``User Database". Then we have compared each set of results using the ``Compare Plot" option for various sets of variations.
\begin{figure*}
\centering
\includegraphics[width=110mm]{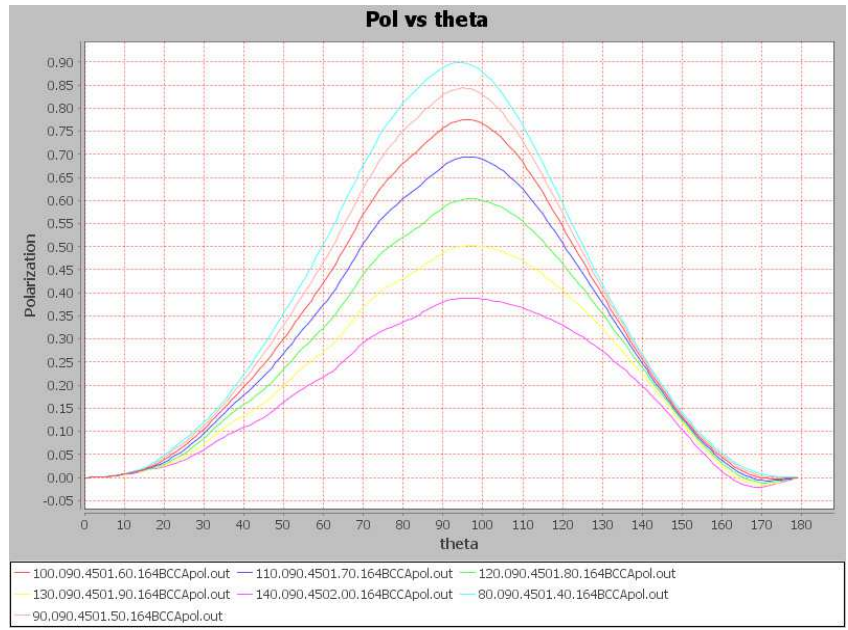}
\caption{Variation in real part of the refractive index.}
\end{figure*}
\begin{figure*}
\centering
\includegraphics[width=110mm]{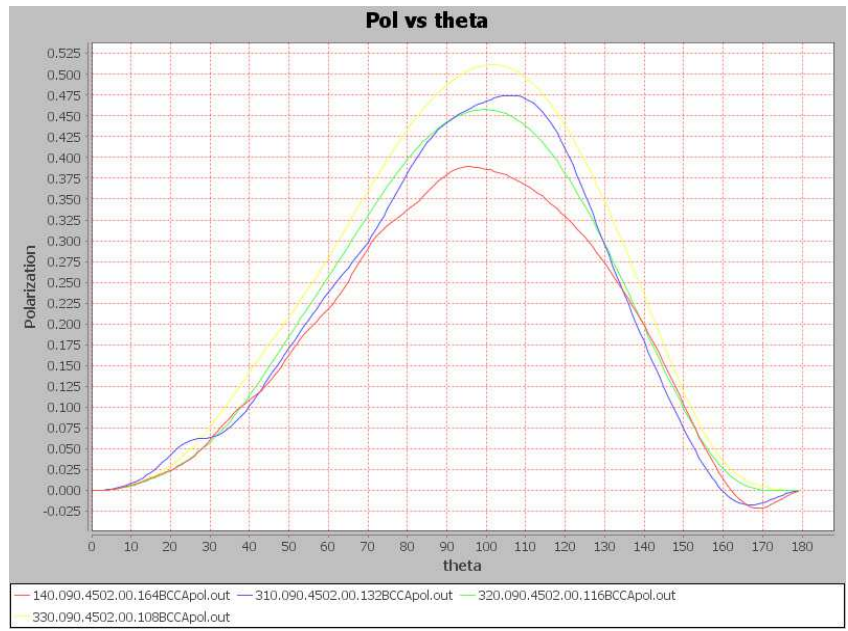}
\caption{Variation in monomer numbers.}
\end{figure*}
\begin{figure*}
\centering
\includegraphics[width=110mm]{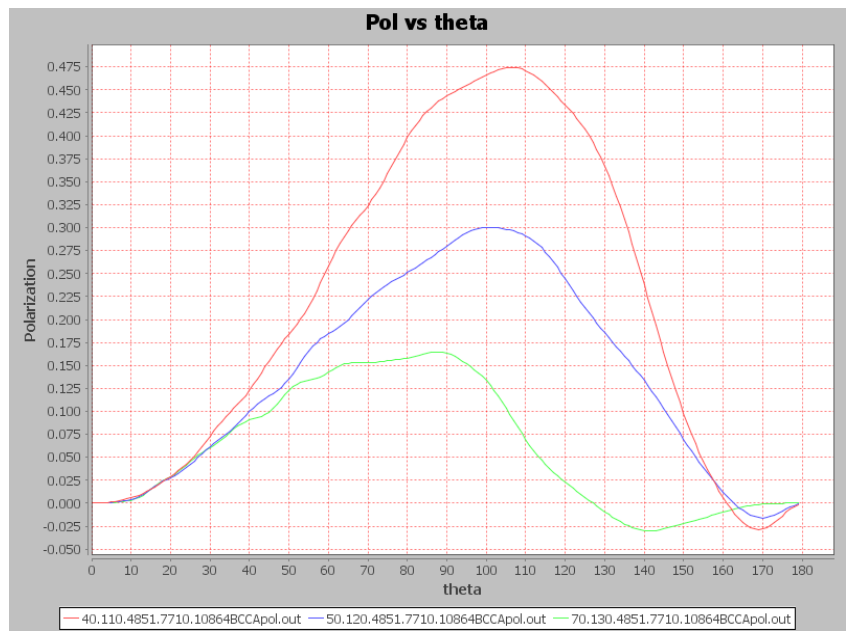}
\caption{Variation in monomer size.}
\end{figure*}
\begin{figure*}
\centering
\includegraphics[width=110mm]{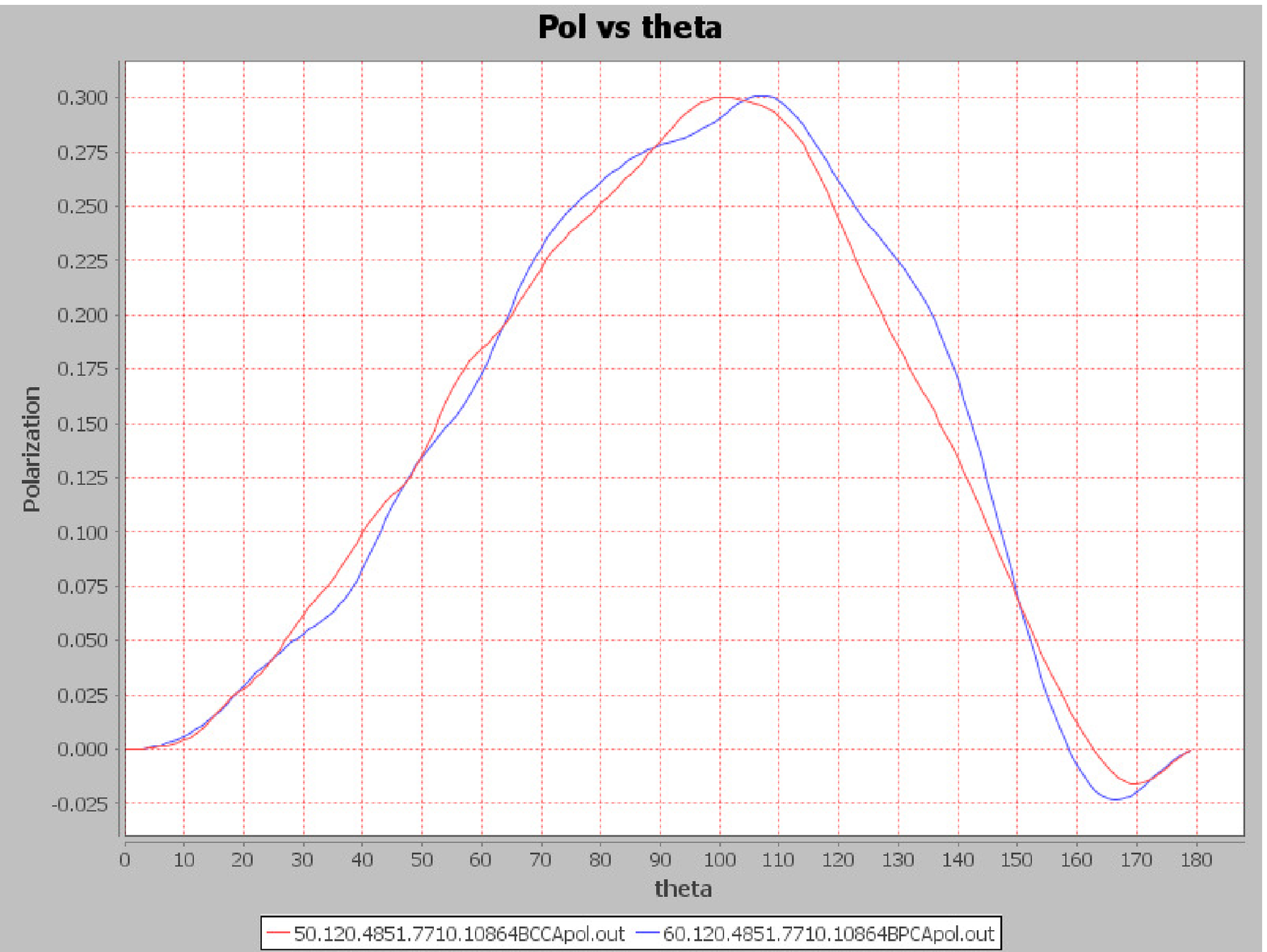}
\caption{Variation in aggregate type.}
\end{figure*}
\subsection{Variation in real part of the refractive index :}
The `Compare Plot' graph in Fig-14 shows the numerical results of polarization as a function of scattering angle. The number of monomers is kept constant at N = 64, the monomer size is fixed at $a_m$ = 0.09 $\mu{m}$, the imaginary refractive index k = 0.1, while the real part of the refractive index is varied from n = 1.4 to n = 2.0 with intervals of 0.1. The figure shows that with increasing number of real refractive index the peak of polarization decreases in the positive axis and increases in the negative axis.

\subsection{Variation in monomer numbers:}
The monomer size is fixed at $a_m$ = 0.09 $\mu{m}$ , imaginary refractive index k = 0.1, the real part of the refractive index is fixed at 2.0, while the monomer number N is varied from N = 08 to N = 64. Fig-15 shows that on increasing the monomer number the polarization value decreases and the peak shifts towards the lower scattering angle.

\subsection{Variation in monomer size:}
In this case we have revisited the works of Das et. al {\cite{Das 2008a}, using JaSTA. We have studied the influence of monomer size on the linear polarization of amorphous olivine by varying the  monomer size {$a_m$} from 0.11 to 0.13 with an interval of 0.01. The number of monomers is fixed at N = 64 whereas the real and imaginary parts of the refractive indices are fixed at n = 1.771 and k = 0.108 respectively ($\lambda$ = 0.485 $\mu{m}$) . The Fig-16 shows that the polarization value decreases and shifts towards the lower scattering angle with increasing monomer size. There is also an emergence of negative polarization for larger monomer size.

\subsection{Variation in aggregate type:}
The monomer number is fixed at N = 64, the real and imaginary parts of the refractive indices are fixed at n = 1.771 and k = 0.108 respectively and the monomer size is considered to be $a_m$ = 0.12 $\mu{m}$. The aggregate type is varied between BCCA and BPCA. Fig-17 shows that there is very little change in the polarization curve. This change may be due to variation of porosity as BCCA is more porous than BPCA.

\begin{figure*}
\centering
\hspace*{-6cm}
\includegraphics[width=120mm]{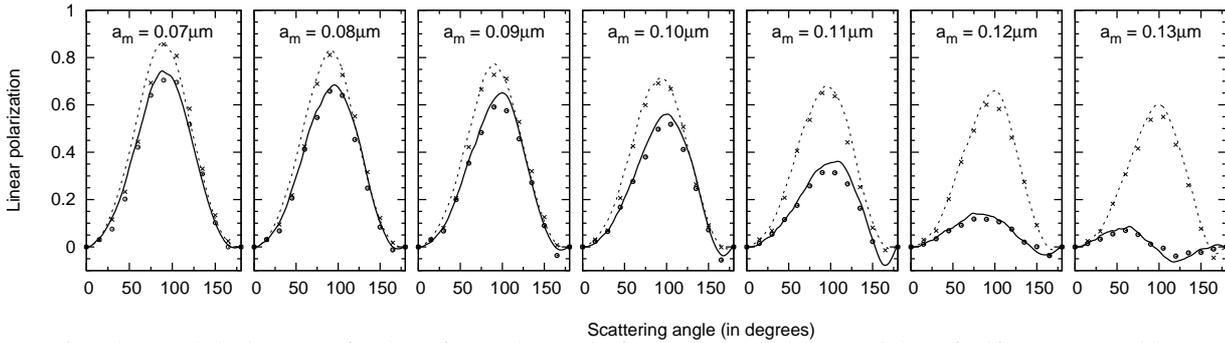}
\caption{Linear polarization as a function of scattering angle for aggregate particles consisting of 128 monomers with monomer size $a_m = 0.07, 0.08, 0.09, 0.10, 0.11, 0.12$ and $0.13\mu m$ with magnesium-rich olivine with the complex refractive index ($m$ = 1.69 + i 0.000104) at a wavelength 0.45$\mu m$ and $m$ = 1.68 + i 0.0000115 at a wavelength 0.60$\mu m$. Solid lines: linear polarization curves at $\lambda$ = 0.45$\mu m$ for BCCA particles; dotted lines: linear polarization curves at $\lambda$ = 0.60$\mu m$ for BCCA particles; open circles: linear polarization curves at $\lambda$ = 0.45$\mu m$ for BPCA particles; crosses: linear polarization curves at $\lambda$ = 0.60$\mu m$ for BPCA particles.}
\end{figure*}

\subsection{Test case}
We provided a test case where the calculations performed by JaSTA are compared against results obtained by Kimura et al. \cite{Kimura 2006}.
They  studied the optical properties of fractal aggregates using the STM code systematically, so we planned to reproduce their results using JaSTA. The calculations are performed with magnesium-rich olivine with the complex refractive index (m = 1.69 + i 0.000104) at $\lambda$ =  0.45 micron and $m$ = 1.68 + i 0.0000115 at $\lambda$ = 0.60$\mu m$ for both BPCA and BCCA structures of 128 monomers (Fig-6 of  Kimura et al. \cite{Kimura 2006}). The results obtained from our computations are shown in Fig. 18. The polarization curves obtained from this work are compared with Kimura et al. and is found to be reasonably good.

\section{Online Support}
For more information and support regarding JaSTA, one can log on to our website \url{http://ausastro.in}. The software package can be downloaded from this website for both Windows OS and Linux OS. We have also included a detailed manual and a video tutorial over there. The link and information on JaSTA are also available in ScattPort\footnote[1]{\url{http://www.scattport.org/index.php/light-scattering-software/t-matrix-codes/list/560-jasta}} , which is a Light Scattering Information Portal for the light scattering community.

\section{Conclusion}
The Java Superposition T-matrix App (JaSTA) has been developed in the Department of Physics, Assam University, Silchar (INDIA) to study the optical properties of cosmic dust aggregates using the STM code. The user can immediately get the output result if it is there in the database of the software making it user friendly. The interactive GUI and database package directly enable a user to model by self-monitoring respective input parameters (viz. wavelength, complex refractive indices, grain size parameter, etc.) to study the related optical properties (viz. extinction, polarization, etc.), of cosmic dust instantly, i.e. with zero computational time. The GUI  of JaSTA has been developed in a very simple manner so that the user can use it very effectively with maximum efficiency and high order accuracy. The current version of this software is developed for the Linux and Windows platform which will be extended for other platforms in future.
We plan to create a large database of  JaSTA in future which will be uploaded frequently in \url{http://ausastro.in}  so that researchers can use it for their work.

\section{Acknowledgment}
This JaSTA is dedicated to the creator of the T-matrix method Peter Waterman. The reviewers of this paper are highly acknowledged for their constructive comments which definitely helped us to improve the quality of the paper. D. Mackowski, K. Fuller and M. Mishchenko are highly acknowledged who made their STM code publicly available. The JaSTA uses ``Double precision superposition codes  for multi-sphere clusters in  random orientations" and is available in \url{http://www.giss.nasa.gov/staff/mmishchenko/t_matrix.html}.  This work is supported by the Department  of Science \& Technology (DST),  Government of India, under SERC-Fast  Track scheme (Dy. No.  SERB/F/1750/2012-13).
\label{}





\bibliographystyle{model1a-num-names}



\end{document}